\documentstyle[aps,prl,multicol,psfig]{revtex} 

\begin{document}

\draft

\title{Metallic Phase and Metal-insulator Transition \\
       in 2d Electronic Systems}

\author{C. Castellani,$^1$ C. DiCastro,$^1$ and P.A. Lee$^2$}

\address{$^1$Istituto di Fisica della Materia e 
Dipartimento di Fisica, Universit\`{a} di Roma ``La
Sapienza,'' Piazzale Aldo Moro 2,
          00185 Roma, Italy \\
$^2$Department of Physics, Massachusetts Institute of Technology,
Cambridge, MA 02139}

\maketitle

\begin{abstract}
The recent experimental observation of a metal-insulator transition in 
two
dimensions prompts a
re-examination of the theory of disordered interacting systems.  We 
argue
that the existing theory
permits the existence of a metallic phase and propose a number of
experiments such as
magnetoconductance and tunnelling in the presence of a parallel field,
which should provide
diagnostic tests as to whether a given experimental system is in fact in
this regime.  We also
comment on a generic flow diagram which predicts a maximum metallic
resistivity.

\end{abstract}

\begin{multicols}{2}

\pagebreak The discovery by Kravchenko {\it et al}. \cite{1,2} of a
metal-insulator transition
(MIT) in a 2 dimensional system (Si-MOSFET) and its confirmation by 
other
workers using different
device designs \cite{3} and materials \cite{4,5} have generated much
excitement because the
conventional wisdom has been that all states are localized in two
dimensions.  Up to now the
discussion of this phenomenon has been based on the scaling theory of
localization of
non-interacting particles, \cite{6,7} even though the possibility of
unusual superconductivity
\cite{8} or spin orbit scattering \cite{9} has also been raised.  On the
other hand, within the scaling theory which includes the
combined effect of interaction and disorder \cite{11,12} a 2d disordered 
system may remain metallic even in the limit of zero temperature \cite{15}. 
In 2d the expansion parameter is the dimensionless resistance per square 
$R_\Box$ defined as $ g = \frac{e^2}{\pi h} R_\Box $.
For weak disorder $(g \ll 1)$ the scaling is towards a metallic state
$(dR_\Box/dT > 0)$ \cite{12,15}.  Furthermore, the theory predicts that a 
magnetic field, via the Zeeman splitting, will drive the system towards 
an insulating state \cite{12,14}.  This is in agreement with
experiment \cite{2}. It is therefore useful to revisit this theory in
light of the recent experimental development.  
One reason why the theory has not received
general acceptance is that the
scaling equations have the peculiar feature that the scaling variables
diverge at some finite value of
the length scale and the theory becomes uncontrolled.  While this is
certainly true in the vicinity
of the MIT where $g \approx 1$, in this paper we reconsider the problem 
of 2d metallic behavior
and argue that for weak disorder the theory remains
under control over a large temperature range, provided the 
renormalization of the energy scale
(relative to the length scale) is taken into account.  In fact this
renormalization allows the
possibility of a metallic state with finite resistance in 2d, in 
contrast to the scaling theory of
localization, which permits only an insulator or a perfect metal ground
state \cite{7}.  We then
study the magnetoresistance and tunneling density of states in the 
presence of a magnetic field, and
point out that these are excellent diagnostic tools to extract key
parameters and to test the
applicability of the theory.  At the end we shall discuss the MIT within
the context of our theory
of the metallic phase and comment on the effects of various symmetry 
breaking perturbations on the scenario we are proposing.
Our main goal is to stimulate experimentalists to further
study the metallic state both in the systems which have been studied up 
to now and possibly in other
promising materials which we will discuss.

We begin by summarizing the results of the scaling theory of interacting
disordered
systems \cite{11,12,15,14,13,10}.  In
addition to the dimensionless resistance $g$, the theory is 
characterized by the coupling constants
$\gamma_2$, $\gamma_c$ and $Z$ which obey the following scaling 
equations:

\begin{eqnarray}
\frac{dg}{dy} & = & g^2  \left[ 1 + 1 + 3 \left( 1 - \frac{1 +
\gamma_2}{\gamma_2}
                    \ln (1 + \gamma_2) \right)  - \gamma_c \right] \\
\frac{d\gamma_2}{dy} & = & g \left[ \frac{1}{2} (1 + \gamma_2)^2 + 
\gamma_c
(1 + 3\gamma_2 +
                          2\gamma_2^2) \right] \\
\frac{dZ}{dy} & = & gZ \left( - \frac{1}{2} + \frac{3}{2} \gamma_2 +
\gamma_c \right) \\
\frac{d\gamma_c}{dy} & = & g \left( \frac{1}{2} + \frac{3}{2}\gamma_2 +
\frac{\gamma_c}{2} -
                           \frac{3}{2}\gamma_2 \gamma_c 
\right) - \gamma_c^2
 \end{eqnarray}

\noindent
where $y = -\ln \lambda$ describes a rescaling of the length scale so 
that
momenta in the range
$\lambda k_0^2 < k^2 < k_0^2$ are integrated out, where $k_0 \approx 
(v_F
\tau)^{-1}$ is the short
distance cut-off with  $\tau$ being the elastic scattering time.  The
parameter $Z$ describes a
rescaling of the energy scale, $Z \gamma_2$ is related to the scattering
amplitude in the triplet
particle-hole channel, while $Z \gamma_c$ is related to the singlet
particle-particle (Cooper
channel) amplitude.  These parameters can be interpreted in the context 
of Fermi liquid
theory \cite{16,17}.  For example, the specific heat linear $T$ 
coefficient is modified by $Z$, so
that $Z$ plays the role of $m^*/m$.  The uniform magnetic susceptibility 
is given by
$\chi_s/\chi_s^0 = Z(1+\gamma_2)$ so that $\gamma_2$ plays the role of 
the Landau parameter
$-A_a^0$.  The key quantity in this theory is the diffusion propagation,
which has a pole of the
form $(Dq^2 - iZ\omega)^{-1}$ where $D$ is related to the conductivity
$\sigma$ (which equals
$R_\Box^{-1}$ in 2d) by $\sigma = \nu_0D$, $\nu_0$ is the bare density 
of states.
In the context of Fermi liquid theory, the diffusion pole can be written 
in the form ($D_{Q}q^2 - i\omega)^{-1}$  where $D_{Q} = D/Z$ has the
interpretation of the
quasiparticle diffusion constant.  Equations (1--3) are derived to 
linear order in $g$ and in the
Cooper amplitude $\gamma_c$ but include all orders in the interaction
amplitude $\gamma_2$.  The
exception is Eq. (4) for $\gamma_c$ where the last term is quadratic in
$\gamma_c$ and independent 
of $g$.
This term renormalizes $\gamma_c$ downwards, so that for $\gamma_c > 0$,
$\gamma_c$ becomes less
important with scaling
and can be neglegted for much of
our subsequent discussions.   The term $1
+ 1$ in Eq. (1) is written in a way 
to remind us that weak localization and singlet
particle-hole channel in the
case of Coulomb interaction give equal contributions to the enhancement 
of
resistivity upon scaling.
The next term is the contribution from the triplet particle-hole 
amplitude
which has the opposite
effect of reducing resistivity.  According to Eqs. (2,3) both $\gamma_2$
and $Z$ grow upon scaling.
In fact, the growth is so rapid that they diverge at a finite scale 
$y_0$,
so that near $y_0$ they behave as 
$\gamma_2 \sim (y_0 - y)^{-1}$ and $Z = (y_0 - y)^{-3}$.
This divergence signals the breakdown of the perturbative scaling 
equations.
Here we want to make
two important points:  (1) the divergence of $Z$ is in fact a necessary
condition for the existence
of a metallic state in 2d; and (2) due to the rapid growth of $Z$ there 
is
a wide range of
temperature where the scaling equations are valid and the system behaves
like a metal.  The key
point is that the growth of $Z$ forces us to perform scaling in an
anisotropic manner in $k$ space
and energy space, a familiar situation in dynamical scaling.  As we
mentioned earlier, the key
quantity is the diffusion pole ($Dq^2 - iZ\omega$).  The scaling 
procedure
then consists of
integrating out the following regions in momentum space and energy space
\cite{14},

\begin{displaymath}
\lambda k_0^2 < k^2 < k_0^2 \,\,\,\, ; \,\,\,\,  \lambda k_0^2  <
\frac{Z}{D} \omega < k_0^2 \,\,\,\, .
\end{displaymath}

\noindent
For $Z$ growing with scaling, the energy or temperature scale decreases
rapidly with scaling, and is
given by

\begin{equation}
T = \lambda Dk_0^2/Z(\lambda) \,\,\,\, .
\end{equation}

\noindent
Strictly speaking, this formula needs further correction when $Z_2 =
Z(1+\gamma_2)$ becomes much
greater than $Z$, because the energy denominator $(Dq^2 - iZ_2 \omega)$
also appears in some
intermediate steps.  However, the qualitative point that the temperature
scale can go all the way
to zero remains valid.  This is important because
 in one parameter scaling, the point has been made that the theory 
scales
to either an
insulator or a perfect metal $(R_\Box \rightarrow 0)$ in 2d, because the
$\beta$ function is always
nonzero. \cite{7}  The diverging $Z$ at $y = y_0$ allows us to escape 
from
this conclusion because in
principle one can reach the point  $y = y_0$ with $g$ 
finite,
so that according to Eq.
(5) the system maintains a finite $R_\Box$ as $T \rightarrow 0$.  

The next question is whether a metallic state can be realized in a 
region of parametric space and
temperature where Eq. (1--3) are valid.  From Eq. (2) and (3), it is
apparent that the effective
expansion parameter in the theory is $g\gamma_2$.   
Then by starting with a sufficiently small $g$, it is
possible to integrate Eqs.
(1--3) until $g\gamma_2$ becomes of order unity.  Since $Z$ diverges as
$(y_0-y)^{-3}$, much faster
than $\gamma_2 \sim (y_0-y)^{-1}$, the scaling can proceed to a rather 
low temperature before
$g\gamma_2 \approx 1$ and the
perturbative equations break down.  By making the assumptions that $g$
approaches a constant
linearly in $(y-y_0)$ we conclude, using Eq. (5), that the low
temperature behavior of the
resistivity is given by $R_\Box(T) = R_0 + c T^{1/3}$ with $c > 0$.
[Notice that  at very low temperature, when $g\gamma_2 \approx 1$, the
assumption that $\gamma_c$ is
negligible is no longer valid and indeed $\gamma_c$ approaches a fixed
point value $\gamma_c^\ast =
1$ for $\gamma_2\rightarrow \infty$.  This would change the behavior of
$Z$, leading to $Z \sim
(y_0-y)^{-\frac{3}{5}}$.  This in turn modifies the temperature 
dependence of $R_\Box=R_0 +
c^\prime T ^{\frac{5}{3}}$ when the regime $\gamma_c \simeq 1$ is 
reached before getting out of the
range of validity of Eqs. (1--4)].
  To summarize, for sufficiently small
$g$, we expect that initially $g$ will exhibit $\ln T$ correction over a
broad temperature range.
If $\gamma_2$ is sufficiently large to begin with, the $\ln T$ 
correction is metallic-like.  If
$\gamma_2$ starts out small, the $\ln$ correction resembles weak
localization, but will change sign below a certain temperature scale 
when $\gamma_2$ has grown
sufficiently to overwhelm the localization term and the singlet
contribution in Eq. (1).  At a
still lower temperature the resistivity drops rapidly, perhaps as 
$T^{1/3}$ (and possibly crossing
over to $T^{\frac{5}{3}}$) before the one loop scaling equation breaks
down \cite{19}.  This qualitative
behavior has been confirmed \cite{20} by numerical integration of Eq.
(1--4).  The point we wish to
emphasize is that these equations predict a metallic behavior down to 
very low temperature in a
region of parameter space where the one loop scaling equations remain
reliable.  Thus the existence
of a metallic state over an experimentally accessible temperture range
should not in itself be a great surprise.

 We have seen that the key ingredient in arriving at a metallic state is 
the existence of a large
$\gamma_2$.  The question is whether $\gamma_2$ can be directly measured
experimentally.  We have
mentioned that the uniform magnetic susceptibility provides a 
measurement of $Z(1 + \gamma_2)$.
However, this is a difficult, though not impossible experiment in a 2
dimensional electron gas \cite{21}.
Instead, we find that magnetoresistance and tunneling in the
presence of a parallel
field provide direct measurements of $\gamma_2$.  A parallel field 
provides a Zeeman splitting of
the spin states which cut off the $S_z = \pm 1$ parts of the triplet
particle-hole channel as well
as the $S_z = 0$ part of the triplet and singlet particle-particle 
channel.
This gives rise to
positive magnetoresistance. The contribution coming from the 
particle-hole channel was calculated in
the weak coupling limit in ref.\cite{22}.  This calculation was later
extended to strong scattering
amplitudes \cite{23}.  Here we further extend this calculation to 
include the effect of the energy
renormalization $Z$.  In analogy with Fermi liquid theory, we expect the
spin splitting of the
quasiparticle to be given by   
$\tilde{\Omega}_s = (1 + \gamma_2) \Omega_s$
where $\Omega_s = g_L\mu_BH$.  Therefore the diffusion pole should be
modified to $(D_{Q}q^2 - i\omega
- i \tilde{\Omega}_s S_z)^{-1}$  for the $S_z = \pm 1$
components of the triplet
particle hole channel.  Inserting this modification into the expression 
for the $S_z = \pm 1$ contribution to the conductivity, we find

\begin{eqnarray}
\delta \sigma(T,H) = 
\frac{ie}{h}^2 \int_{-\infty} ^\infty d\omega \frac{d}{d\omega} \left(
\omega \coth
\frac{\omega}{2kT} \right)  \int \frac{d^2k}{(2\pi)^2} D_{Q}^2k^2  
\nonumber \\
\sum_{S_z = \pm 1}\frac{1} {(D_{Q}k^2 - i\omega - 
i\tilde{\Omega}_sS_z)^2}
\frac{2\gamma_2}{D_{Q}k^2 - i(1+\gamma_2)\omega - i\tilde{\Omega}_sS_z}
\end{eqnarray}

\noindent
The parameters $D$, $Z$ and $\gamma_2$ in this equation are scale
dependent.  Noting that the
contributions for small $H$ are dominated by small $k$ and $\omega$, we
evaluate these parameters at
the scale $\lambda$ given by Eq. (5).  The integrals are then performed
following ref. \cite{22}.
In particular, we find that for small $H$,

\begin{equation}
\sigma(H,T) - \sigma(0,T) = -0.084 \frac{e^2}{\pi h} 
\gamma_2(\gamma_2+1)
\left( \frac{g_L\mu_BH}{kT} \right)^2 \,\,\,\, .
\end{equation}

\noindent
We recover the weak coupling limit by setting $\gamma_2 \rightarrow F/2$
where $F \ll 1$ is the
interaction parameter in ref. \cite{22}.  If we include the Cooper 
channel contribution, we will
find an additional contribution of
$
-0.084 \frac{e^2}{\pi h} \gamma_c(\gamma_2 + 1)^2 (g_L \mu_BH/kT)^2.
$
The above treats the effect of spin splitting only and is appropriate 
for $H$ parallel to the
plane.  For perpendicular field we have, in addition to Eq. (7), the 
usual weak localization
negative magnetoresistance.  In this case there is an additional
contribution proportional to
$\gamma_c$ but now the orbital field scale given by $\Omega_H = 4DeH/c$
also enters as a cut-off and
the magnetic field dependence from this term is more complicated.  Since 
in the weak coupling regime we
expect $\gamma_c$ to
scale to weak coupling, we shall concentrate on Eq. (7).  The main 
point is that the
quadratic in $H$ term in parallel field
magnetoresistance provides a measurement of the parameter $\gamma_2$.  
It will be very interesting
to see if this parameter is indeed large in the metallic MOSFET samples 
and whether it increases
with decreasing temperature.  The available data are not systematic 
enough to answer these
questions in the metallic regime.  Most of the experiments on
magnetoresistance are close to the
MIT and for fields with $\Omega_s \geq kT$.  Qualitatively, the 
(positive) magnetoresistance
increases as one moves away from the MIT \cite{3}.  This is in agreement
with our expectation that
$\gamma_2$ should consistently increase in order to establish a metallic 
phase.

Another way to measure $\gamma_2$ is by tunnelling experiment.  It was
pointed out that the
tunnelling density of states exhibit additional structure between the
energy scales of the bare
spin splitting $g_L\mu_BH$ and the enhanced spin splitting due to
interaction effects \cite{23}.
Following the Fermi liquid analogy, this second energy scale should be
given by $\tilde{\Omega}_s$. In
particular, in 2d the derivative of the tunnelling density of states has
logarithmic singularities at $\omega = g_L\mu_BH$ and $\omega = (1 +
\gamma_2) g_L\mu_BH$.  Thus
tunnelling gives a direct measurement of $\gamma_2$.  Recently a new
technique has been developed
to tunnel into a 2d electron gas \cite{24}.  It will be very interesting 
to apply it to the new metallic samples.

As the field is increased, we expect a cross-over to the strong Zeeman
splitting universality
class.  The detailed cross-over is complicated, but the high field limit 
is one of the few fixed
points which is controlled.  The system always scales to an insulator, 
and in the weak disorder
limit, a universal logarithmic temperature dependence was predicted 
\cite{14}:
$\sigma(T) = \sigma_0 + (e^2/\pi h) (2-2 \ln 2) \ln (T\tau)$.
As far as we know, this prediction has never been tested.  The new 
MOSFET samples offer an ideal
testing ground for this prediction.

Up to now we have limited our discussion to the weak disorder case, when
Eq. (1--4) remains valid.
We now comment on the possibility of the existence of a nontrivial fixed
point if somehow the
scaling equations can be extended to strong coupling.  In ref. \cite{18}
the 2 loop contribution to the scaling equations was evaluated under the 
assumption of $\gamma_2 \gg 1$ but for small $g\gamma_2$. 
The two loop scaling equations or ref. \cite{18} indeed exhibit a
non-trivial fixed point. From this fixed point two separatrices originate 
ending at $\gamma_2=0$ and $\gamma_2=\infty$.  Since the interesting
part of the flow diagram is 
not in the weak coupling regime, the 
scaling equations and the
details of the flow cannot be trusted.  Nevertheless, the structure of 
the flow may be generic. 
Here we wish to make some general comments.  If the initial $\gamma_2$ 
is not too large, the system
exhibits a metal-to-insulator transition.  An interesting feature of 
this flow is that on the
metallic side of the separatrix the system reaches infinite $\gamma_2$ 
and $Z$ at a finite scale
$\lambda$ as in one loop order.  
Thus the discussion we gave earlier in this paper still holds
and a metallic state with
finite $R_\Box$ is possible at $T=0$.  In fact, the metallic state in 
the low $T$ limit exhibits a maximum metallic
resistivity given by $\rho_{M}=(\pi h/e^2)g_{M}$, 
where $g_M$ is the value of $g$ 
on the separatrix at $\gamma_2=\infty$. This $g_M$ is in
general smaller than the
value $g^*$ at the fixed point. Experimentally $\rho^{*}=(\pi h/e^2)g^{*}$ 
is determined as the resistance which
separates the metallic and insulating states at higher temperature.  
This feature seems to be
consistent with currently available data.  For example, the data of ref.
\cite{1} yields $\rho_M \approx
0.1 \frac{h}{e^2}$ and $\rho^* \approx 2 \frac{h}{e^2}$.

The scaling behavior near the MIT will be controlled both by the 
existence
of a fixed point at finite $g^{*}$ and $\gamma_{2}^{*}$ and by the 
runaway towards $g\simeq g_{M}$ and $\gamma_{2}=\infty$.
Then one can show that $R_\Box = \tilde{\rho}(T/(\delta n)^{\nu z})$
where $\delta n$ is the deviation from the critical density and the
critical indices $\nu$ and $z$ are determined by the fixed point.
$\tilde{\rho}$ is a scaling function and 
according to the previous discussion $\tilde{\rho}(\infty)=(\pi h/e^2)g^{*}$ 
and $\tilde{\rho}(0)=(\pi h/e^2)g_{M}$.

Besides the magnetic field, other
symmetry breaking perturbations have relevant  effects on 
our picture of the 2d metallic phase. Spin flip scattering by magnetic
impurities will cause a crossover to a low $T$ insulating phase. The 
effect of
spin orbit (SO) scattering is more intriguing. In $d=2$, intrinsic SO 
coupling or SO scattering by impurities only affects the out of plane component 
of the spin  \cite{25}. In this case the one loop 
equations \cite{26}
still lead to a diverging behavior of the ($S_{z}=0$) triplet 
amplitude and a
metallic phase at low T. We suggest that the above discussion on 
the MIT applies in this case even though the 2d SO could result into a 
different universality class. 
A much more dramatic effect on our theory of the metallic phase is the 
SO scattering deriving from possible asymmetry of the confining potential 
since it is equivalent to a 3d SO coupling and cutoff all triplets \cite{27}. 
If this
coupling is sizeable, the theory predicts an insulating behavior at zero
temperature \cite{28}, at least in the limit in which the SO band 
splitting is less than the inverse elastic scattering time.
In our opinion, evidences of 2d or 3d SO are still
lacking.

The scenario we outlined in this paper has the
advantage of permitting a
metallic state in 2d and therefore a metal-insulator transition. 
However, given the uncertainties of the strong 
coupling theory, a good
strategy is to approach the MIT from the metallic side and try to gain a
thorough understanding of
the metallic state.  This motivates us to propose magnetic 
susceptibility, magnetoresistance and
tunnelling experiments as ways to directly measure the key parameters of
the theory $\gamma_2$ and
$Z$.  We also worked out the qualitative behavior of the temperature
dependence of the resistivity,
in a regime where the theory is valid.  Here our results do not compare
favorably with experiments.
The data of ref. \cite{1} and ref. \cite{4} have been fitted to the form
$\rho(T) = \rho_0 +
\exp(-T_0/T)$.  This is very different from the $\ln T$ dependence 
followed
by a low temperature
power law that we predict.  Furthermore, the parameter $T_0$ appears to
scale with the Fermi energy
which is relatively small in these low density systems.  Thus the
possibility remains that some
physics on the scale of the Fermi energy is playing the dominant role 
and the data are far from
the low energy scaling regime we considered here.  We believe these
questions can be addressed by
more detailed studies of the metallic state along the lines suggested in
this paper.  Yet another
possible research direction to confirm the theory here presented 
is to study 2d systems where $\gamma_2$ is
expected to be large to begin
with, such as almost ferromagnetic metallic thin films.  Examples are  
weak ferromagnets such as MnSi or TiBe$_2$, if the ferromagnetism can be
suppressed by alloying \cite{29,30}.

{\it Acknowledgements} We thank Michael Ma and Olav Sylju\aa sen for helpful
discussions.  PAL
acknowledges the support of NSF under Grant No. DMR--9523361.

\end{multicols}


\begin{references}
\bibitem{19}It is worth noticing that the two loop equations of ref.
\cite{18} would lead to a low temperature behavior 
$R_\Box(T) - R_\Box(0) \propto
            T^{\frac{1}{2}}\exp\left( -C/T^{1/2}\right)$.  
However since at the same time $g\gamma_2 \gg 1$, 
there is no control on this result.
\end{references}
\end{document}